# Regular and Anomalous Motion of Individual Magnetic Quincke Rollers Under Rotating Magnetic Field

Zoran M. Cenev,[1,2,†] Ville S.I. Havu,[1] and Jaakko V.I. Timonen[1,‡]

[1]Department of Applied Physics, Aalto University, 02150 Espoo, Finland
[2]Department of Mechanical and Production Engineering, Aarhus University, 8200 Aarhus, Denmark.

**ABSTRACT**. We report the motion of individual magnetic Quincke rollers composed of silica particles doped with superparamagnetic iron oxide nanoparticles, whose activity arises from the coupling between Quincke rolling and an externally applied rotating magnetic field. We applied a clockwise (CW) rotating magnetic field of magnitude approximately 11 mT and rotational frequencies ranging from 0.2 to 2.75 Hz. At low frequencies, the dominant mode of motion is a CW helical trajectory. Circular trajectories emerge as a limiting case of this helical motion, in which lateral translation vanishes and the particle traces overlapping closed loops in the xy-plane. At higher frequencies, a second regular mode becomes prevalent, characterized by helical wavy trajectories in which the particle follows a CW helical path with a spatially varying curvature. Under specific conditions, however, we observe the unexpected emergence of anomalous counterclockwise (CCW) trajectories, in which individual particles roll in a direction opposite to that of the applied CW rotating magnetic field. A theoretical model incorporating electrostatic interactions, far-field hydrodynamic coupling, and a magnetic dipole approximation indicates that the anomalous behavior results from the interplay among the magnitude and orientation of the initial magnetic dipole moment, the frequency of the rotating magnetic field, and the magnitude of the initial translational velocity. Together, these factors determine the likelihood of a particle exhibiting regular or anomalous rotational motion.

## I. INTRODUCTION

Active matter is a system in out-of-thermodynamic equilibrium, consisting either of living organisms that convert energy into motion or of inorganic particles whose motion is induced by external stimuli. These constituents of active matter systems can interact with each other, leading to a complex collective behavior across many scales, spanning from human crowds [1,2], flocking birds [3], school of fish [4], insect swarms [5], bacterial suspensions [6–8], Janus particles in a hydroxyl peroxide bath [9], magnetic particles in time-varying magnetic fields [10–13], dumbbell-shaped colloids in AC electric fields [14], and other systems as reviewed in [15–19]. An individual constituent in an active matter system may be simply called an active particle [20] or an active Brownian particle [21].

A particular case of active matter is Quincke rollers. Quincke rollers are dielectric spheres immersed in a weakly conducting fluid in between two capacitor plates that roll in a random direction once subjected to a sufficient homogeneous electric field, reported more than a century ago by Quincke himself [22]. The presence of the dielectric spheres in the weakly conducting fluid increases the effective conductivity, which in turn facilitates the ions' migration [23] from one electrode plate to the other. These rollers can exhibit self-organization to achieve coherent motion [24]. Beyond the spherical shape of the rollers, the rolling has also been investigated for ellipsoidal particles [25], self-assembled dumbbells and trimers [26], and liquid droplets [27–29].

Recently, we demonstrated the behavior of magnetic Quincke rollers (MQRs) when subjected to a uniform, directional magnetic field; the random component in their rolling direction is strongly suppressed, leading to rolling predominantly perpendicular to the magnetic field. [30] More recently, Fitzgerald et al. 2025 [31] showed that under linear magnetic fields $\geq 20$ mT, MQRs execute stable paths parallel to the uniform B field, demonstrating simultaneous right-angled rolling. We note that the dynamics of individual magnetic Quincke rollers in a homogeneous and continuously rotating magnetic field (RMF) remain unexplored.

Here, we present experimental observations and theoretical analysis of individual magnetic Quincke rollers subjected to a uniform continuously rotating magnetic field with a fixed magnitude of 11.2 mT. For a clockwise (CW) uniform rotating magnetic field, the individual MQRs dominantly follow the magnetic field and rarely rotate in the opposite (CCW) direction. We refer to the former as the "regular" motion and the latter as the "anomalous" motion. By employing electrostatic interactions, far-field hydrodynamic coupling, and a magnetic dipole approximation, our theoretical model attributes the magnitude and orientation of the initial magnetic dipole moment, the frequency of the rotating magnetic field, and the magnitude of the initial translational velocity as the determining factors that dictate whether an individual MQR will likely exhibit regular or anomalous rotational behavior.

[†] Corresponding author: zoran.cenev@mpe.au.dk
[‡] Corresponding author: jaakko.timonen@aalto.fi

## II. EXPERIMENTAL RESULTS

Similar to our previous work [30], we use commercially available spherical $SiO_2$ superparamagnetic microparticles (Microparticles GmbH) with a diameter $\phi \approx 21.6$ μm. The magnetic moment arises from the embedded iron oxide nanoparticles within the microparticles. ~1 μL of microparticle suspension was immersed in 10 to 20 μL of slightly conductive liquid medium ($\sigma \approx 10^{-8}$ S/m) [27] made of n-dodecane with 150 mM of sodium bis (2-ethylhexyl) sulfosuccinate (AOT). The dispersion was inserted in a Hele-Shaw cell comprising two transparent indium tin oxide-coated glass slides with an average gap of ~35 μm or ~45 μm. The particles were subjected to a constant voltage of 36 or 37 V, yielding a homogeneous electric field $E_0$ from 0.81 to 1.05 V/μm.

The magnetic field originated from two (1" × 1") cylindrical NdFeB permanent magnets facing each other with different polarity fixed at 132 mm. This configuration produced a nearly uniform magnetic field of 11.2 mT at the midpoint, as measured by a Hall sensor, and maintained a nearly uniform field within the workspace. The magnets were mounted on a brushed motor with a planetary gear head. We varied the motor frequency, which in turn translates to the frequency of the rotating magnetic field, from 0.2 to 2.75 Hz. The applied rotation was in the CW direction throughout all experiments. In our experiments, we first turn on the rotating magnetic field and then switch on the electric field. Additional details on the experimental setup and notes on the experiments are provided in Appendix A.

Figures 1(a-i) and 1(b-i) illustrate the two distinct motion modes exhibited by individual magnetic Quincke rollers in the homogeneous rotating magnetic field. In the regular motion mode (Fig. 1(a) and Video S1), the particles rotate in the same (clockwise) direction as the applied rotating field. A very common motion mode is a CW helical trajectory, as shown in Fig. 1(a-ii). Circular trajectories emerge as a special limiting case of the helical motion in which lateral translation vanishes and the particle traces overlapping closed loops in the plane, as shown in FIG 1(a-iii). Another regular motion mode, which is more common for higher frequencies (2 to 2.75 Hz), is the helical wavy trajectories. Here, the particle's trajectory is CW helical with a changing sign of curvature (i.e., wavy), as shown in FIG 1(a-iv). Other sample trajectories for eight different frequencies of the clockwise rotating magnetic field are provided in Figure S1 in the Supporting Information [32].

In contrast, the anomalous motion mode is characterized by magnetic Quincke rollers rotating in the opposite CCW direction, despite the magnetic field rotating in the CW direction. We refer to this behavior as *anomalous* due to its rarity. This motion mode was observed at RMF frequencies of 0.2, 0.46, and 0.86 Hz, as shown in Figs. 1(b-ii), 1(b-iii), and 1(b-iv), respectively. The processed video data is provided in Video S2, see also notes in Supporting Information.

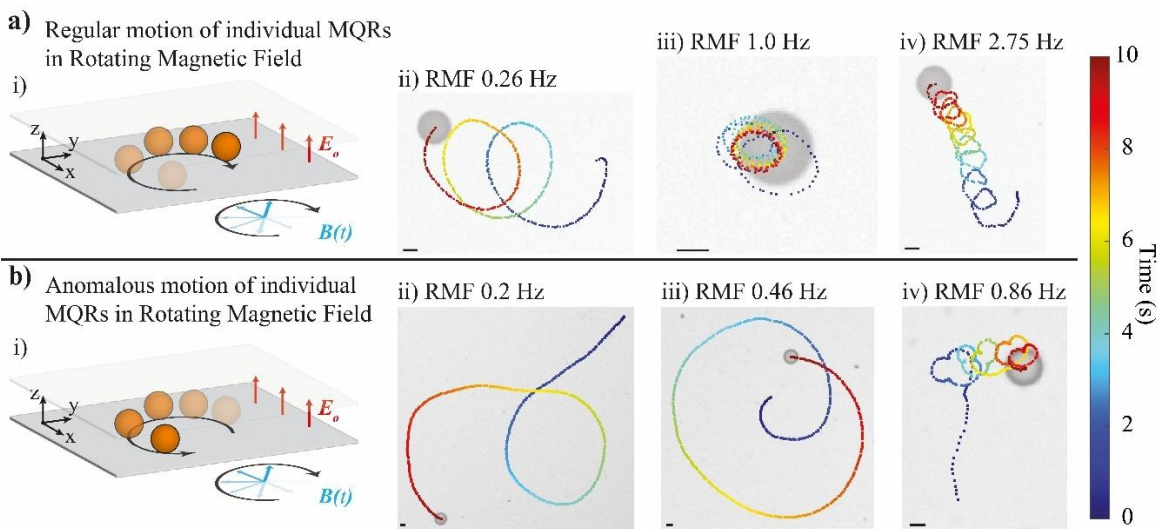

**FIG 1**. Illustration and experimental evidence of the motion modes of individual Magnetic Quincke Rollers (MQRs) in a homogeneous rotating magnetic field of 11 mT. **a)** Regular motion: Three different individual MQRs roll in a clockwise (CW) direction when the magnetic field also rotates CW. *(i)* illustration; *(ii)* 0.26 Hz, *(iii)* 1.0 Hz, and *(iv)* 2.75 Hz rotating-field frequencies. **b)** Anomalous motion: Three different individual MQRs rolling counterclockwise (CCW) despite the magnetic field rotating CW. *(i)* illustration; *(ii)* 0.2 Hz, *(iii)* 0.46 Hz, and *(iv)* 0.86 Hz rotating-field frequencies. Scale bars: 10 μm.

FIG 2 presents a quantitative analysis of the anomalous motion modes observed experimentally for individual magnetic Quincke rollers in Fig.1(b-ii to iv). Time-resolved particle position and velocity data are shown for the three representative rotating magnetic field frequencies of 0.2 Hz [Fig.2(a)], 0.46 Hz [Fig.2(b)], and 0.86 Hz [Fig.2(c)]. The corresponding average particle velocities are 132.2 µm/s, 126.3 µm/s, and 40.7 µm/s, computed from the smoothed data. In the case of 0.86 Hz, shown in Fig. 2(c), the particle velocity exhibits recurrent local minima, occurring 12 times in total, with the velocity reaching zero in 10 of these events. In these cases, the particle stops its movement and changes direction, as is evident in the trajectory in Fig. 1(b-iv) and Video S2. Also, the average velocity is lower in this case due to the particle stopping and reaccelerating.

### III. THEORETICAL MODEL

From a theoretical perspective, an important advantage of the individual MQRs subjected to uniform magnetic field is that the dominant interaction mechanisms can be unambiguously identified. We derive the equations of motion for individual MQRs interacting via electrostatic and far-field hydrodynamic interactions, [21] coupled with a magnetic dipole approximation. The angular velocity $\boldsymbol{\Omega}$ of an individual MQR under uniform magnetic field in matrix form is given as:

$$\begin{pmatrix} \Omega_x \\ \Omega_y \\ \Omega_z \end{pmatrix} = \begin{pmatrix} \mu_r & 0 & 0 \\ 0 & \mu_r & 0 \\ 0 & 0 & \mu_\perp \end{pmatrix} \begin{pmatrix} T_x^e + T_x^m \\ T_y^e + T_y^m \\ T_z^m \end{pmatrix} \quad (1)$$

Where $\mu_r$ and $\mu_\perp$ denote the in-plane rolling friction and sliding friction coefficients, respectively. The quantities $T_{x,y}^e$ and $T_{x,y,z}^m$ represent the components of the electric and magnetic torques along the corresponding Cartesian axes.

The translational velocity $\boldsymbol{v}$ can be expressed in matrix form as:

$$\frac{1}{a}\begin{pmatrix} v_x \\ v_y \end{pmatrix} = \tilde{\mu}_t \begin{pmatrix} 0 & 1 \\ -1 & 0 \end{pmatrix} \begin{pmatrix} T_x^e + T_x^m \\ T_y^e + T_y^m \end{pmatrix}. \quad (2)$$

Where $a$ is the particle diameter, and $\tilde{\mu}_t$ is the tangential mobility coefficient. A detailed derivation of these expressions is provided in Appendix B.

### IV. SIMULATION RESULTS

In our simulations, we define a random direction and magnitude (20…200 µm/s) for the initial velocity, as well as a random orientation and magnitude ($3\times10^{-13}$…$3\times10^{-11}$ Am$^2$) of the initial permanent magnetic moment of the particle. The simulation time was set to 10 seconds, the same as in the experiments. For each RMF frequency, 10,000 rollers with random initial values were simulated. The other constant parameters are given in Table S1.

Figure 3 presents numerically simulated trajectories of individual magnetic Quincke rollers subjected to a rotating magnetic field at frequencies corresponding to those shown experimentally in Fig. 1. In the regular regime [Fig. 3(a)], the simulations reproduce circular trajectories at 0.26 Hz and 1.0 Hz, while at 2.75 Hz the motion transitions to a wavy helical trajectory characterized by periodic modulation of the curvature. The helical nature of the trajectories is not well captured at low frequencies (0.2 to 0.6 Hz) but becomes apparent at mid-range frequencies (0.86 to 1.5 Hz) and even more pronounced at high frequencies (2 to 2.75 Hz), as shown in Figs. S2-S4, whereas the wavy characteristics can be seen in all frequency ranges.

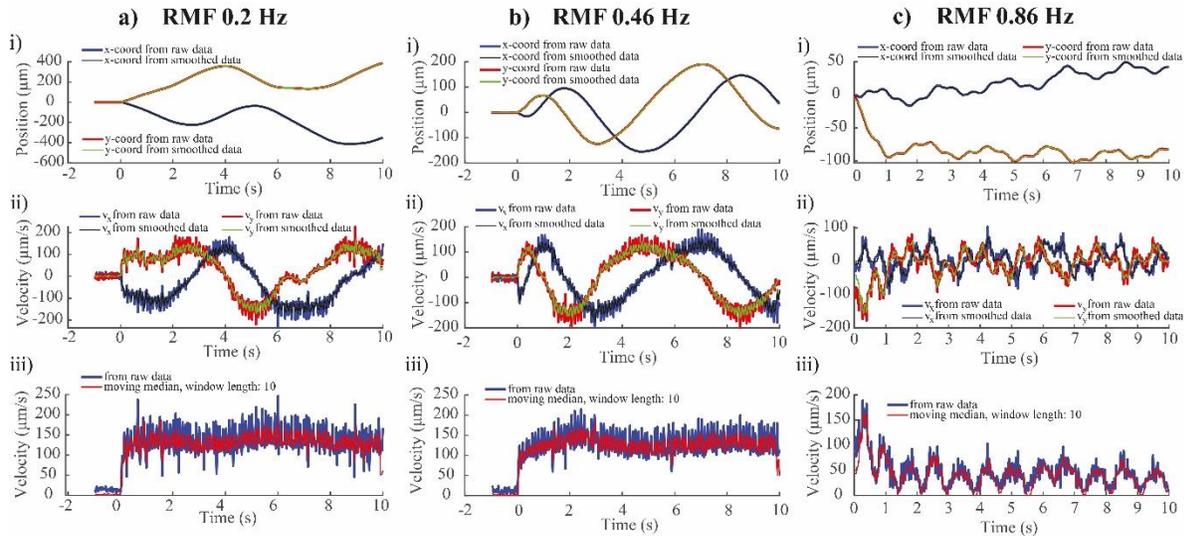

**FIG 2.** Analysis of the experimental data of the anomalous motion modes of individual Magnetic Quincke Rollers presented in Fig.1(b) for Rotating Magnetic Field frequencies of **a)** 0.26, **b)** 0.46, and **c)** 0.86 Hz. Panels (i) show the particle position as a function of time; panels (ii) show the x- and y-velocity components as a function of time, and panels (iii) show the planar velocity as a function of time. The electric field is switched on at $t = 0$ s.

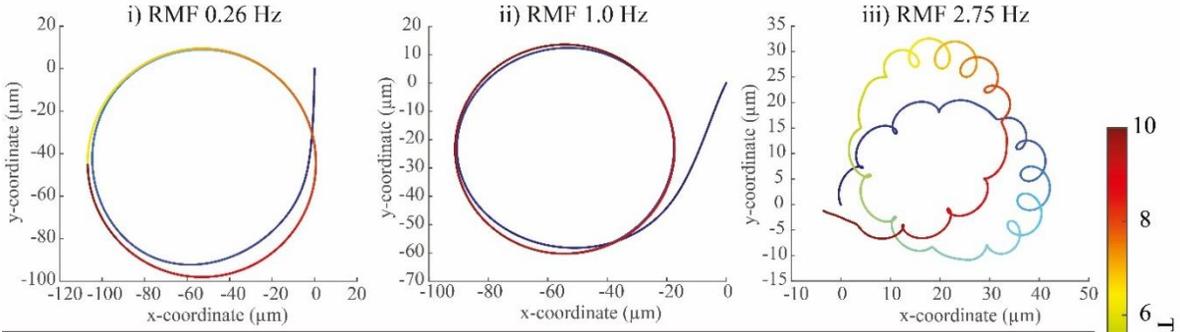

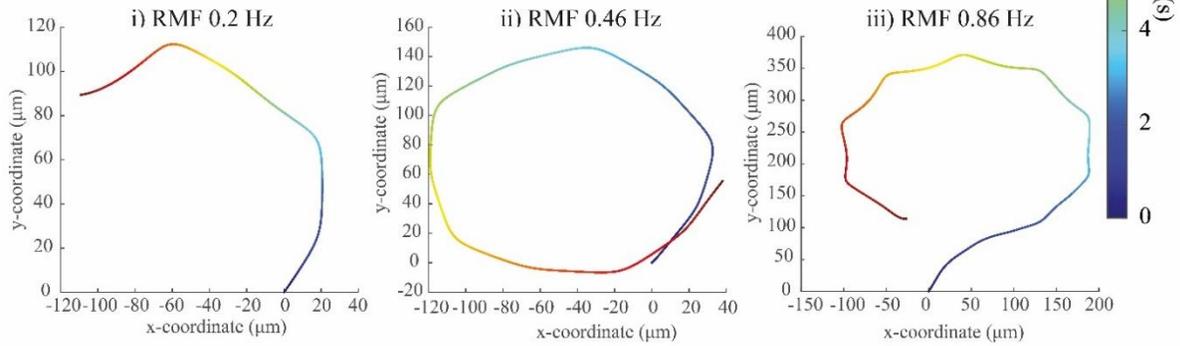

**FIG 3.** Numerically simulated trajectories of a single magnetic Quincke roller under a rotating magnetic field. **(a)** Regular motion modes at RMF frequencies of (i) 0.26 Hz, (ii) 1.0 Hz, and (iii) 2.75 Hz. **(b)** Anomalous motion modes at RMF frequencies of (i) 0.20 Hz, (ii) 0.46 Hz, and (iii) 0.86 Hz.

The simulations also capture the anomalous regime [Fig. 3(b)] with 3,220 anomalous cases out of 100,000 simulated. At an RMF frequency of 0.2 Hz, the particle follows an incomplete loop trajectory [Fig. 3(b-i)]. Although this trajectory does not close into a loop as observed experimentally, it nevertheless reproduces the key characteristics of anomalous rotation. At 0.46 Hz, the simulated trajectory closes into a single loop [Fig. 3(b-ii)], in agreement with the experimental observation [Fig. 1(b-iii)]. At 0.86 Hz, the simulated motion becomes wavy but unclosed [Fig. 3(b-iii)]. While this case captures the irregular nature of the anomalous dynamics, it does not fully reproduce the helical features observed experimentally [Fig. 1(b-iv)].

FIG 4 summarizes the statistics of anomalous motion for individual MQRs under a rotating magnetic field. To exclude trajectories that are effectively straight, the analysis applies a threshold on the mean turning angle.

Figure 4(a) reports the number of anomalous trajectories with a mean turning angle exceeding 0.01°. Results without this threshold are shown in Fig. S5.

The model predicts the occurrence of anomalous motion over the full range of investigated frequencies, from 0.2 to 2.75 Hz [Fig. 4(b)], consistent with experimentally observed frequencies. According to our model, the probability peaks at 1.5 Hz, with approximately 1000 anomalous trajectories, corresponding to a 10% occurrence rate, and is significantly lower at other frequencies.

FIG 4(c) shows that anomalous motion is most likely for permanent magnetic moments in the range $0.1–3 \times 10^{-11}$ A m$^2$. Fig. 4(d) presents the distribution of anomalous cases as a function of the initial magnetic-moment orientation, displayed as a color map in azimuthal and elevation angles. The likelihood of anomalous motion is increased for elevation angles between $-90°$ and $-45°$ and between $45°$ and $90°$.

FIG 4(e) shows that the probability of anomalous motion is increasing with the increase in the magnitude of the velocity of an anomalously moving MQR. Finally, Fig. 4(f) shows that the direction of the velocity vector does not play a role in whether an anomalous motion would occur or not since its distribution is rather uniform.

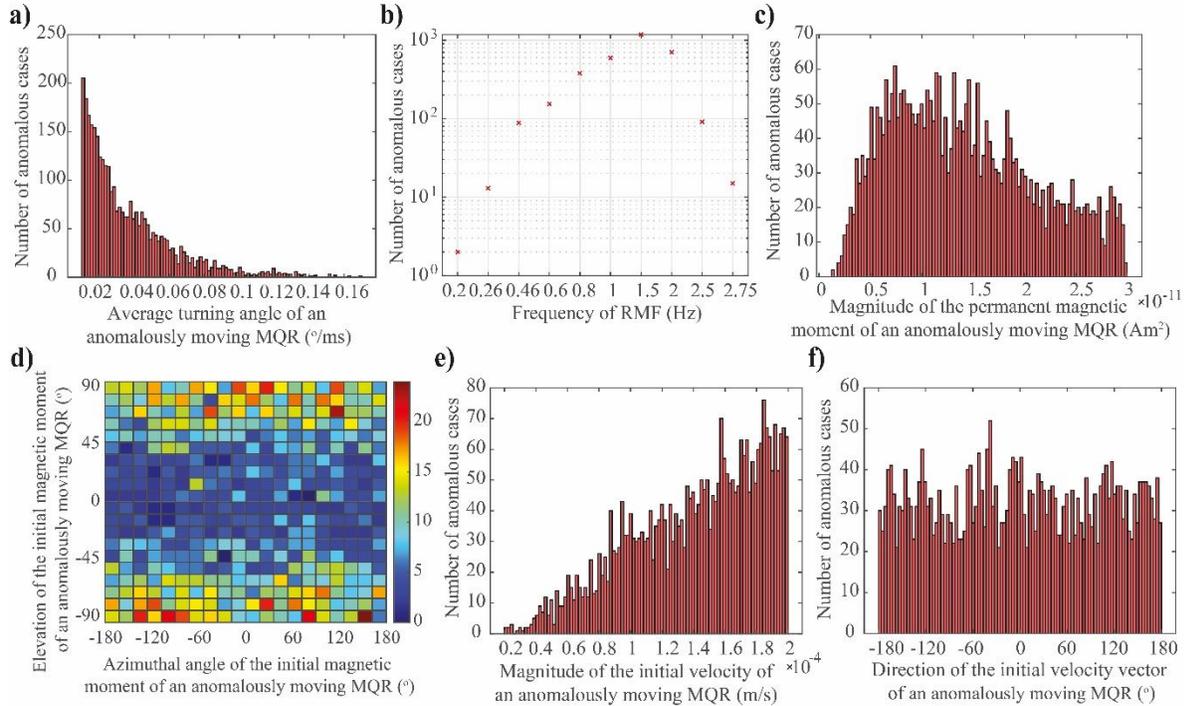

**FIG 4.** Statistics of anomalous motion for individual magnetic Quincke rollers under a rotating magnetic field. Number of simulated anomalous trajectories as a function of (a) average turning angle, (b) frequency of rotating magnetic field, (c) magnitude of the permanent magnetic moment, (d) initial magnetic-moment orientation, parameterized by azimuthal and elevation angles, (e) magnitude of the initial velocity of an anomalously moving MQR, and (f) direction of the of the initial velocity of an anomalously moving MQR. A total of $10^4$ simulations were performed for each frequency, yielding $10^5$ simulated cases overall.

## V. DISCUSSION

In Figs. 2(a-b), one can see that for anomalous motion, the components of position and velocity are not sinusoidal, indicating that the trajectory is not circular. This contrasts with the regular motion, where almost circular trajectories have been observed, see Fig. S1. Similar differences are seen in the simulated trajectories, where the regular motion is often purely circular, whereas the anomalous motion resembles a polygon (see Fig. 3).

Our theoretical model and simulations are built on top of the theory presented by Bricard et al. [24] with the addition of the magnetic field coupled to the magnetic moment of the particle similar to Fitzgerald et al. 2025 [31]. The simulations reproduce both the regular and anomalous motion of the rollers. The coupling between the magnetic moment and the rotating magnetic field gives rise to $\Omega_x$ and $\Omega_z$, respectively, of the angular velocity $\Omega$, which consequently result in changes in the x and y components of the translational velocity $v_x$ and $v_y$, respectively. Rotation of the particle in turn changes the magnetic moment, giving rise to complex motion. Our theoretical model can also explain the particle motion in a homogeneous magnetic field, as we already reported in [30].

The probability of a roller executing anomalous motion depends on the initial conditions, according to the simulations. If the initial magnetic moment is aligned with the $xy$-plane, the roller is likely to be a regular one. However, if there is a considerable out-of-plane component of the initial magnetic moment, the probability of observing an anomalous roller is increased 5-20-fold depending on the orientation of the magnetic moment, see Fig. 4(d). Further, the probability for an anomalous roller is increased along with the increase in the initial velocity of the roller. However, it is independent of the initial direction of the roller, see Fig. 4(e-f).

The fraction of the predicted anomalous rollers is ~3%, which is in agreement with previous experimental results for right-angle turning of the magnetic field [30]. The simulations show that the behavior of the rollers is stochastic in the sense that, in the experiments, one cannot select parameters to execute a given trajectory. However, changing the frequency of the RMF will change the distributions of the rollers, meaning certain modes of motion are more probable for certain frequencies.

The initial distribution of the magnetic moments in the experiments is probably not uniform, but in-plane alignment is preferred since the magnetic field is switched on prior to the electric field, and the rollers have some time to align their magnetic moment with the magnetic field. In such cases, anomalous motion is unlikely to occur.

## VI. CONCLUSIONS

Here, we showed the behavior of the individual MQRs in a homogeneous continuously rotating magnetic field. Motion is dominantly regular helical, in special cases circular, and at high frequencies (2 to 2.75 Hz), the helical trajectory becomes wavy-helical. The motion can also be anomalous in some rare cases. Due to the rarity of the anomalous motion, we have observed only three cases presented in Fig. 1(b). Despite the limited sample size, which prevents a systematic classification of trajectories, the observations unambiguously demonstrate the existence of anomalous motion.

## ACKNOWLEDGMENTS

ZMC acknowledges the funding support from the formerly known Academy of Finland (now Research Council of Finland) under Grant 342268. JVIT acknowledges the funding support from European Research Council grant 803937 and Academy of Finland Center of Excellence Program grant 346112. Contributions: ZMC developed the experimental setup, conducted experiments, observed, and identified the anomalous behavior of the magnetic Quincke rollers, analyzed data, and wrote the manuscript. VH and ZMC developed and implemented the theoretical model; JVIT conceived the research on magnetic Quincke rollers, contributed to the modifications of the experimental setup, provided lab equipment, and supervised the research.

## APPENDIX A: DETAILS ON EXPERIMENTAL SETUP

The experimental setup was previously reported in [30]. Briefly, the setup involved the use of two cylindrical NdFeB permanent magnets (K&J Magnetics, Inc., DX0X0-N52) fixed on a dovetail rail (Thorlabs, Inc., XT66SD-250) approximately 132 mm apart, facing each other with different polarity. This configuration generated a nearly uniform magnetic field of 11.2 mT at the midpoint between the magnets, which was measured using a gaussmeter (SENIS AG, Hall Probe).

To rotate the magnetic field uniformly, a planetary geared direct current (DC) motor (Micro Motors SRL, E192.24.25) was employed, with the dovetail attached to it. The motor was controlled using a motor driver (Seeed Technology Co., MD13S) and a microcontroller board (BCMI, Arduino Uno) connected to a computer. Power for the motor and the driver was supplied by a benchtop DC power supply (Multicomp Pro, MP710086) set at 24 V. The rotation speed of the dovetail, and consequently the magnetic field direction, was regulated using pulse-width modulation (PWM) or manual current adjustment to the power supply.

Prior to the actual experiments, the rotational speed of the motor corresponding to different current inputs was determined experimentally. This was achieved by employing a hall sensor (Honeywell, SS495A/SS495A1) positioned below the dovetail rail. The hall sensor was connected to an analog-to-digital/digital-to-analog instrumentation device (NI Inc., myDAQ) to record the motor's rotational frequency for a duration of 100 or 500 seconds.

Imaging of the sample was accomplished using a custom-built microscope comprising an infinity-corrected objective lens (Mitutoyo Inc., 5×/0.14), a tube lens (Thorlabs Inc.), a CMOS camera (Basler AG, acA2440-75um), and coaxial illumination provided by a white LED (Thorlabs Inc., MCWHLP1) with the aid of a beam splitter (Thorlabs Inc., BSW10R).

To minimize mechanical vibrations, the driving system was isolated from the imaging system by setting them up on separate adjacent tables. The sample cell was firmly held in place at the midpoint between the rotating magnets using optomechanical components (Thorlabs Inc.).

The microparticle-AOT-Dodecane suspension was stored in a low-humidity chamber (relative humidity ≈ 5%) to reduce water diffusion within the AOT-Dodecane suspension and consequently result in particle charging.

## APPENDIX B: DETAILED DERIVATION OF THE THEORETICAL MODEL

Here, we consider a perfect dielectric sphere with radius $a$ located at $\boldsymbol{r} = 0$ (own coordinate system), rolling in a weakly conductive Newtonian fluid with angular velocity $\boldsymbol{\Omega}$ and translational velocity $\boldsymbol{v}$. The electric permittivity of the particle and the weakly conducting fluid are denoted with $\epsilon_p$ and $\epsilon_l$, respectively. Similarly, the respective electric conductivities are $\sigma_p$ and $\sigma_l$. The mismatch of the properties is characterized with the following ratios $S_\epsilon = \epsilon_p/\epsilon_l$, and $R_\sigma = \sigma_p/\sigma_l$. The solid particle is assumed to be impermeable. The supplied electric field is uniform and time-invariant, along the vertical direction, such that $\boldsymbol{E}_0 = E_0\hat{\boldsymbol{z}}$, where $\hat{\boldsymbol{z}}$ is the unit vector along the vertical z-direction.

### 1. Induced electric field polarization from uniform electric field

Even though our microparticles contain magnetic nanoparticles, we assume perfect dielectric surface (shell) of the microparticles such that it enables electric charges to interplay at the particle-liquid interface when there is the electric field present in the workspace. The Quincke rotation comes into play when the applied electric field becomes greater than the threshold electric field $E_0 > E_Q$ causing the charge relaxation time of the particle to be greater than the charge relaxation time of the liquid, $\tau_p > \tau_l$ such that $\tau_{p,l} = \epsilon_{p,l}/\sigma_{p,l}$, [23] giving rise to an in-plane electric torque which acts on the spherical particle according to $\boldsymbol{T}_\parallel^e = \frac{\epsilon_l}{\epsilon_0} \boldsymbol{P} \times \boldsymbol{E}_0$, where $\boldsymbol{P}$ is the electric polarization of the particle. The individual components of $\boldsymbol{T}_\parallel^e$ then become $T_x^e = \frac{\epsilon_l}{\epsilon_0} P_y \cdot E_0$, and $T_y^e = -\frac{\epsilon_l}{\epsilon_0} P_x \cdot E_0$. In the absence of a magnetic field the in-plane electric torque is directly related to the in-plane angular velocity as [24]

$$\boldsymbol{\Omega}_\parallel = \mu_r \cdot \boldsymbol{T}_\parallel^e \qquad (A1)$$

with a rolling friction factor $\mu_r = 8\pi a^3 \eta$.

### 2. Magnetic dipole approximation of a magnetic Quincke roller in a uniform magnetic field

The magnetic dipole of a Quincke roller is oriented as $\boldsymbol{m} = m_x \hat{\boldsymbol{x}} + m_y \hat{\boldsymbol{y}} + m_z \hat{\boldsymbol{z}}$ in an external uniform magnetic field $\boldsymbol{B} = B_x \hat{\boldsymbol{x}} + B_y \hat{\boldsymbol{y}} + B_z \hat{\boldsymbol{z}}$ subjected to the workspace in the magnetically tagged particles. The torque on the magnetic dipole, or the magnetic torque, therefore, becomes $\boldsymbol{T}^m = \boldsymbol{m} \times \boldsymbol{B}$. Assuming $B_z = 0$ [T], the components of the magnetic torque acting on the magnetic Quincke roller are:

$$\begin{aligned} T_x^m &= -m_z B_y \\ T_y^m &= m_z B_x \\ T_z^m &= m_x B_y - m_y B_x \end{aligned} \qquad (A2)$$

For a magnetic Quincke roller, the angular velocity, in addition to the electric torque, arises also from the magnetic torque. Hence

$$\boldsymbol{\Omega}_\parallel = \mu_r (\boldsymbol{T}_\parallel^e + \boldsymbol{T}_\parallel^m). \qquad (A3)$$

The z-component for the angular velocity only contains the magnetic component since the electric polarization acts along the xy plane. Rearranging equations S1-S3, one obtains the angular velocity of an individual magnetic Quincke roller as

$$\begin{aligned} \Omega_x &= \mu_r(\frac{\epsilon_l}{\epsilon_0} P_y E_0 - m_z B_y) \\ \Omega_y &= \mu_r(-\frac{\epsilon_l}{\epsilon_0} P_x E_0 + m_z B_x) \\ \Omega_z &= \mu_\perp(m_x B_y - m_y B_x) \end{aligned} \qquad (A4)$$

With $\mu_\perp$ being the sliding friction coefficient.
Equation (S4) in matrix notation becomes:

$$\begin{pmatrix} \Omega_x \\ \Omega_y \\ \Omega_z \end{pmatrix} = \begin{pmatrix} \mu_r & 0 & 0 \\ 0 & \mu_r & 0 \\ 0 & 0 & \mu_\perp \end{pmatrix} \begin{pmatrix} T_x^e + T_x^m \\ T_y^e + T_y^m \\ T_z^m \end{pmatrix} \qquad (A5)$$

### 3. Velocity of a Quincke roller in a uniform electric field

The velocity of an individual Quincke roller in a uniform electric field has been already derived in [24] as

$$\boldsymbol{v}_{qr} = -\frac{\epsilon_l}{\epsilon_0} a \widetilde{\mu}_t E_0 \boldsymbol{P}_\parallel^\sigma \qquad (A6)$$

where $\widetilde{\mu}_t$ is the tangential mobility factor and $\boldsymbol{P}_\parallel^\sigma$ is the dynamic in-plane polarization [23]. For the case of individual MQR, the translational velocity $\boldsymbol{v}$ besides the electric polarization contains also the contribution from the magnetic torques, hence

$$\begin{aligned} v_x &= a \widetilde{\mu}_t (T_y^e + T_y^m) \\ v_y &= -a \widetilde{\mu}_t (T_x^e + T_x^m) \end{aligned} \qquad (A7)$$

or

$$\begin{aligned} v_x &= -a \widetilde{\mu}_t \left( \frac{\epsilon_l}{\epsilon_0} P_x \cdot E_0 - m_z B_x \right) \\ v_y &= -a \widetilde{\mu}_t \left( \frac{\epsilon_l}{\epsilon_0} P_y \cdot E_0 - m_z B_y \right) \end{aligned} \qquad (A8)$$

which in matrix notation becomes:

$$\frac{1}{a} \begin{pmatrix} v_x \\ v_y \end{pmatrix} = \widetilde{\mu}_t \begin{pmatrix} 0 & 1 \\ -1 & 0 \end{pmatrix} \begin{pmatrix} T_x^e + T_x^m \\ T_y^e + T_y^m \end{pmatrix} \qquad (A9)$$

During the rolling motion also the direction of the velocity and the magnetic moment of an individual roller change. In the implementation of the above equations, they are updated at every timestep $dt$ with a rotation around the axis defined by the angular velocity $\Omega$. The amount of rotation is given by the angle $d\theta = |\Omega| dt$. After the rotation the polarization is projected in-plane and renormalized, but the magnetic moment can rotate freely giving rise to complex motion.

[1] D. Helbing, I. Farkas, and T. Vicsek, Simulating dynamical features of escape panic, Nature **407**, 487 (2000).

[2] N. Bain and D. Bartolo, Dynamic response and hydrodynamics of polarized crowds, Science (1979). **363**, 46 (2019).

[3] A. Cavagna and I. Giardina, Bird flocks as condensed matter, Annu. Rev. Condens. Matter Phys. **5**, 183 (2014).

[4] E. Crosato, L. Jiang, V. Lecheval, J. T. Lizier, X. R. Wang, P. Tichit, G. Theraulaz, and M. Prokopenko, Informative and misinformative interactions in a school of fish, Swarm Intelligence **12**, 283 (2018).

[5] M. Sinhuber and N. T. Ouellette, Phase Coexistence in Insect Swarms, Phys. Rev. Lett. **119**, 1 (2017).

[6] T. Danino, O. Mondragón-Palomino, L. Tsimring, and J. Hasty, A synchronized quorum of genetic clocks, Nature **463**, 326 (2010).

[7] A. Sokolov and I. S. Aranson, Physical properties of collective motion in suspensions of bacteria, Phys. Rev. Lett. **109**, 1 (2012).

[8] S. Liu, S. Shankar, M. C. Marchetti, and Y. Wu, Viscoelastic control of spatiotemporal order in bacterial active matter, Nature **590**, 80 (2021).

[9] J. R. Howse, R. A. L. Jones, A. J. Ryan, T. Gough, R. Vafabakhsh, and R. Golestanian, Self-Motile Colloidal Particles: From Directed Propulsion to Random Walk, Phys. Rev. Lett. **99**, 8 (2007).

[10] A. Kaiser, A. Snezhko, and I. S. Aranson, Flocking ferromagnetic colloids, Sci. Adv. **3**, 1 (2017).

[11] A. Snezhko and I. S. Aranson, Magnetic manipulation of self-assembled colloidal asters, Nat. Mater. **10**, 698 (2011).

[12] A. Snezhko, I. S. Aranson, and W.-K. Kwok, Surface Wave Assisted Self-Assembly of Multidomain Magnetic Structures, Phys. Rev. Lett. **96**, 078701 (2006).

[13] A. Snezhko and I. S. Aranson, Magnetic manipulation of self-assembled colloidal asters., Nat. Mater. **10**, 698 (2011).

[14] A. R. Sprenger, M. A. Fernandez-Rodriguez, L. Alvarez, L. Isa, R. Wittkowski, and H. Löwen, Active Brownian Motion with Orientation-Dependent Motility: Theory and Experiments, Langmuir **36**, 7066 (2020).

[15] S. Ramaswamy, The mechanics and statistics of active matter, Annu. Rev. Condens. Matter Phys. **1**, 323 (2010).

[16] M. C. Marchetti, J. F. Joanny, S. Ramaswamy, T. B. Liverpool, J. Prost, M. Rao, and R. A. Simha, Hydrodynamics of soft active matter, Rev. Mod. Phys. **85**, 1143 (2013).

[17] D. Needleman and Z. Dogic, Active matter at the interface between materials science and cell biology, Nat. Rev. Mater. **2**, 17048 (2017).

[18] C. J. O. Reichhardt and C. Reichhardt, Ratchet effects in active matter systems, Annu. Rev. Condens. Matter Phys. **8**, 51 (2017).

[19] S. Shankar, A. Souslov, M. J. Bowick, M. C. Marchetti, and V. Vitelli, Topological active matter, Nature Reviews Physics **4**, 380 (2022).

[20] F. Schmidt, H. Šípová-Jungová, M. Käll, A. Würger, and G. Volpe, Non-equilibrium properties of an active nanoparticle in a harmonic potential, Nat. Commun. **12**, (2021).

[21] P. Romanczuk, M. Bär, W. Ebeling, B. Lindner, and L. Schimansky-Geier, Active Brownian particles: From individual to collective stochastic dynamics: From individual to collective stochastic dynamics, European Physical Journal: Special Topics **202**, 1 (2012).

[22] G. H. Quincke, Ueber Rotationen im constanten electrischen Felde, Annalen Der Physik Und Chemie **295**, 417 (1896).

[23] N. Pannacci, L. Lobry, and E. Lemaire, How insulating particles increase the conductivity of a suspension, Phys. Rev. Lett. **99**, 2 (2007).

[24] A. Bricard, J. B. Caussin, N. Desreumaux, O. Dauchot, and D. Bartolo, Emergence of macroscopic directed motion in populations of motile colloids, Nature **503**, 95 (2013).

[25] Q. Brosseau, G. Hickey, and P. M. Vlahovska, Electrohydrodynamic Quincke rotation of a prolate ellipsoid, Phys. Rev. Fluids **2**, 1 (2017).

[26] A. Mauleon-Amieva, M. P. Allen, T. B. Liverpool, and C. P. Royall, Dynamics and interactions of Quincke roller clusters: From orbits and flips to excited states, Sci. Adv. **9**, eadf5144 (2023).

[27] G. Raju, N. Kyriakopoulos, and J. V. I. Timonen, Diversity of non-equilibrium patterns and emergence of activity in confined electrohydrodynamically driven liquids, Sci. Adv. **7**, 1 (2021).

[28] P. M. Vlahovska, Electrohydrodynamics of drops and vesicles, Annu. Rev. Fluid Mech. **51**, 305 (2019).

[29] Q. Dong and A. Sau, Unsteady electrorotation of a viscous drop in a uniform electric field, Physics of Fluids **35**, (2023).

[30] R. Reyes Garza, N. Kyriakopoulos, Z. M. Cenev, C. Rigoni, and J. V. I. Timonen, Magnetic Quincke rollers with tunable single-

particle dynamics and collective states, Sci. Adv. **9**, 2 (2023).

[31] E. Fitzgerald, C. Clavaud, D. Das, I. C. D. Lenton, and S. R. Waitukaitis, Rolling at right angles: magnetic anisotropy enables dual-anisotropic active matter, (2025).

[32] Z. M. Cenev, V. S. I. Havu, and J. V. I. Timonen, Regular and Anomalous Motion of Individual Magnetic Quincke Rollers Under Rotating Magnetic Field-Supporting Information, n.d.